# Nanoscale polarization manipulation and encryption based on dielectric metasurfaces

*Ruizhe Zhao[1], Lingling Huang[1,4], Chengchun Tang[2], Junjie Li[2],*
*Xiaowei Li[3], Yongtian Wang[1], Thomas Zentgraf[4]*

[1] School of Optics and Photonics, Beijing Institute of Technology, Beijing, 100081, China
[2] Institute of Physics, The Chinese Academy of Sciences, Beijing, 100191, China
[3] Laser Micro/Nano-Fabrication Laboratory, School of Mechanical Engineering, Beijing Institute of Technology, Beijing 100081, China
[4] Department of Physics, University of Paderborn, Warburger Straße 100, D-33098 Paderborn, Germany

**Abstract:** Manipulating the polarization of light is highly desired for versatile applications ranging from super resolution, optical trapping, to particle acceleration. The enormous freedom in metasurface design motivates the implementation of polarization control in ultrathin and compact optical systems. However, the majority of proposed strategies based on metasurfaces have been demonstrated only a spatially homogeneous polarization generation, while less attention has been devoted to spatially variant inhomogeneous vector beams. Here, we demonstrate a novel method for generating arbitrary radial and azimuthal polarization beams with high efficiencies of up to 80% by utilizing transmission-type dielectric metasurfaces. Polarization conversion metasurfaces are suitable candidates for the implementation of polarization encryption, which we demonstrate by encoding a hidden image into the spatial polarization distribution. In addition, we show that the image pattern can be modified by appropriate polarization selection of the transmitted light. Such a method may provide a practical technique for a variety of applications such as imaging, encryption and anti-counterfeiting.

**Main:**

Metasurfaces as a novel class of flat optical elements have attracted enormous interest due to their ability to arbitrarily tailor the fundamental properties of the electromagnetic wavefront within ultrashort distance by engineering the interface between two materials [1-3]. A wide variety of applications have been proposed based on plasmonic metasurfaces composed of metallic antennas [4-6]. However, the low efficiency caused by high metal absorption loss hinders transmission-type single-layer metallic metasurfaces for further practical applications. Alternatively, dielectric metasurfaces consisting of a monolayer of high refractive index nanoresonators with low loss have been demonstrated to improve the efficiency by utilizing both electric and magnetic resonances [7-9]. By judiciously designing the geometric shape, size and orientation of the dielectric nanoresonators, the wavefront can be controlled at will. Dielectric metasurfaces have successfully achieved novel applications of beam shaping [10], anomalous refraction [11, 12], nonlinear optics [13-16], circular dichroism [17, 18], spin-hall effect of light [19-21], and holography [22, 23].

While many applications of metasurfaces are based on pure phase modulation, some efforts have been made to realize a full control of the light properties by using a polarization modulation [24-28]. By utilizing the superposition of two circular components in the reflection direction, reflection-type metasurfaces have been demonstrated recently for polarization control [24, 25]. Another strategy uses cascaded metasurfaces for generating broadband polarization conversion [26]. Based on high-contrast dielectric elliptical nanoposts, a transmission-type metasurface was proposed to simultaneously control phase and polarization by using elliptical high-index nanoposts with an efficiency of more than 90% [27]. However, a majority of the proposed strategies based on metasurfaces have demonstrated only a spatially homogeneous polarization generation, while less attentions have been devoted to spatially variant inhomogeneous vector beams and polarization encryption schemes.

Cylindrical vector (CV) beam are a group of vector beams which are solutions of the vector wave equation that obey cylindrical symmetry and possess spatially inhomogeneous states of polarizations [29]. Two important subcategories of CV beams are radial polarization and azimuthal polarization beams. Compared to the beams with spatially homogeneous polarization, the radial polarization beams can be focused into a tighter spot due to the strong longitudinal electric field at the focal plane, while the azimuthal polarization beams can be focused into a doughnut shape intensity pattern [30]. Because of these specific focusing properties, the CV beams have been applied in particle acceleration [31], optical trapping [32, 33], super-resolution [34, 35] and mechanical fabrication processing [36]. Driven by these peculiar applications, many methods have been proposed to generate CV beams, such as using Sagnac interferometer [37], Brewster prism with polarization selectivity [38], spatial light modulators (SLMs) combined with a 4-$f$ system [39] and segmented subwavelength gratings that work as a quarter wave plate [40]. However, the above-mentioned methods required either advanced electro-optical systems (like the SLM) or complex experimental setups, which are not suitable for compact optical systems due to their use of bulky optical elements. In addition, if the working wavelength moves to the visible or ultraviolet regime, the requirement of characteristic subwavelength feature sizes becomes more challenging to achieve due to the difficulty of nanofabrication, which may limit practical applications. In order to overcome the shortcomings that exist in conventional methods, dielectric metasurfaces can provide an excellent alternative scheme for nanoscale polarization manipulation.

In this letter, we propose and experimentally demonstrate an ultrathin and compact optical element based on dielectric metasurfaces for generating high-order radial and azimuthal polarization beams. The metasurfaces consist of an array of silicon nanofins with identical geometric parameters but spatial variant orientations. Each nanofin acts as a local half-wave plate, whereas its long and short axis can be considered as the fast and slow axis, respectively. A schematic illustration of the generation process of high-order radial and azimuthal polarization beams (with the order $P=4$) is shown in **Figure 1**. The dielectric metasurface can convert $x$ and $y$ linearly polarized light into radial and azimuthal polarization beams by suitable arrangement of the orientation of each nanofin. When the polarization angle of the incident linearly polarized light varies, the transmitted beam will become a superposition of both radial and azimuthal polarization beams which results in a new type of CV beams. To further validate our proposed approach, we also successfully demonstrate an image encryption with inhomogeneous polarization distribution. Such image pattern can be modified by selecting the suitable polarization state of the transmitted light. Our proposed method provides a simple strategy for manipulating the polarization of light and may result in a variety of applications for compact integrated optical systems such as vector beam generation and polarization encryption.

Our designed dielectric metasurfaces consist of amorphous silicon nanofins on top of a glass substrate (**Figure 2**a). The high-index dielectric nanofin can be considered as a tiny waveguide which has a corresponding effective refractive index ($n_{eff}$) for two orthogonal linearly polarized beams (corresponding to the eigenmodes of the waveguide) [41]. Hence, the rectangular cross-section of the nanofin results in form birefringence with different effective refractive indices along each axis due to the geometric size. The effective index and therefore the phase delay can be tailored by adjusting the length $L$ and width $W$ of nanofin which are shown in **Figure S1.** Therefore, a phase delay will occur for linearly polarized light since the decomposed polarization components along the long or the short axis of the nanofin propagate at different velocities. Hence, the output light will change its polarization due to such phase delay between the electric field components. In order to achieve the conversion from linear polarization to radial and azimuthal polarization, the spatial variant half-wave plates are designed to achieve high transmission and effective polarization conversion. When a linearly polarized light is propagating through a half-wave plate, the transmitted light can be calculated by Jones matrix method as follows [42]:

$$\begin{bmatrix} \cos 2\theta & \sin 2\theta \\ \sin 2\theta & -\cos 2\theta \end{bmatrix} \begin{bmatrix} \cos \alpha \\ \sin \alpha \end{bmatrix} = \begin{bmatrix} \cos(2\theta - \alpha) \\ \sin(2\theta - \alpha) \end{bmatrix} \quad (1)$$

In Equation1, the Jones matrix denotes a half-wave plate with its fast axis rotated by an orientation $\theta$ with respect to the *x*-axis, and $\alpha$ is the polarization angle of incident linearly polarized light. Obviously, the polarization angle of the transmitted beam is rotated by $2\theta$ when illuminating with *x* linearly polarized light ($\alpha=0°$), as shown in **Figure 2**b. For *y* linearly polarized light ($\alpha=90°$), the polarization of the transmitted light will change to the orthogonal direction compared to the former case. Therefore, to achieve full control of the spatially inhomogeneous states of polarization, it is sufficient to rotate the orientation angle of each nanofin locally. In such way, high-order radial and azimuthal polarization beams can be generated.

To achieve such functionality with a dielectric metasurface made of silicon, we carry out a 2D parameter optimization by using a rigorous coupled wave analysis method (RCWA). The length *L* and width *W* of nanofin are swept in the range of 80 nm to 200 nm and 40 nm to 150 nm, respectively, while maintaining the height *H* at 500 nm and the period size *S* at 240 nm. The values of *H* and *S* are carefully chosen to guarantee the phase delay between orthogonal transmission polarizations can be equal to $\pi$ by considering the suitable length and width of the nanofin. In addition, we restricted the range of the geometric parameters to values that are feasible for our nanofabrication. For the simulation, the nanofin is placed on a glass substrate ($n_{SiO2}$=1.5). The wavelength of incident light is fixed at 780 nm and we used for the corresponding refractive index of amorphous silicon a value of $n_{Si}$=3.8502+0.010918*i*. The 2D parameter optimization results are listed in the Supplementary Materials. As an example, we extract and plot the amplitude and phase difference of the transmission coefficients $t_{xx}$ and $t_{yy}$ for *W* ranging from 40 nm to 150 nm with *L*=180 nm (**Figure 2**c). For the width of *W*=90 nm of the nanofin the phase difference between $\varphi_x$ for *x* polarized light and $\varphi_y$ for *y* polarized light equals to $\pi$. Simultaneously, the transmission amplitudes for the two transmission coefficients

$t_{xx}$ and $t_{yy}$ are both over 90%. Therefore, such silicon nanofins can work as high-efficiency half-wave plates. To confirm the result of the RCWA calculations we simulated the phase distribution $\varphi_x$ and $\varphi_y$ for periodically arranged nanofins for the illumination with two orthogonal linearly polarized beams by using a finite element method. From the results, it is obvious that the phase of the transmitted waves is shifted by half a wavelength with respect to each other (**Figure 2**d). We simulated normalized magnetic energy densities of a single nanofin (180×90×500 nm$^3$, $L \times W \times H$) with periodic boundary conditions when illuminating with linearly x-polarized light, as shown in **Figure 2**e. It's obvious that the light is mainly confined inside the high-refractive-index nanofin. Meanwhile, the effect caused by neighboring nanofins with different orientation angles can be neglected [27]. The polarization distribution of the transmitted light can be tailored pixel by pixel based on a metasurface that consists of an array of silicon nanofins with identical geometric parameters but spatial variant orientation angles.

To achieve arbitrary high-order CV beams, we arrange the local orientation of each nanofin depending on the polarization of the desired beam. The polarization distribution Φ of CV beams in polar coordinates is determined by:

$$\Phi(r,\phi) = P \times \phi + \phi_0 \tag{2}$$

where $(r, \phi)$ are the polar coordinates. $P$ and $\phi_0$ indicate the order and initial angle of the polarization of the CV beams. In order to simplify the design of metasurfaces and generate radial polarization beams by illuminating with an $x$ linearly polarized beam, the initial angle of polarization $\phi_0$ is set to zero and the orientation angle $\theta$ of the nanofin can be calculated by $\theta = \Phi/2$ according to Equation 2. By using the finite difference time domain method, we simulate three dielectric metasurface samples for generating high-order radial and azimuthal polarization beams with $P$=3, 4, 5. The metasurfaces are all composed of 41×41 nanofins on top of the glass substrate, as shown in **Figures 3**a-c. The incident polarizations are along $x$- and $y$-axis at a wavelength of $\lambda$=780 nm. To better demonstrate the polarization state in each point, we extract the complex amplitude of electromagnetic field and calculate the ellipticity $X$ and orientation of the polarization ellipse $\psi$ according to Equation 3.

$$\begin{cases} X = \tan \chi = \mp \dfrac{b}{a} \quad (-1 \leq X \leq 1) \\ \tan 2\psi = \dfrac{2E_{ax}E_{ay}\cos\delta}{E_{ax}^2 - E_{ay}^2} \quad (0 \leq \psi \leq \pi) \end{cases} \quad (3)$$

whereas $a$ and $b$ represent the length of the long and the short axis of each polarization ellipse, which can be calculated by the complex amplitude of electromagnetic field. The positive and minus signs correspond to right and left elliptical polarization. $\chi$ denotes the ellipticity angle and the ellipticity $X$ is defined as the ratio of two axes of the polarization ellipse. Meanwhile, $E_{ax}$ and $E_{ay}$ indicate the amplitude of the $x$ and $y$ component of the electromagnetic field ($E_x$ and $E_y$) and the phase difference between $E_x$ and $E_y$ is expressed by $\delta$. **Figures 3**d-i show that radial and azimuthal polarization beams are generated by orthogonal linearly polarized light. Our simulations show that the polarization of the transmitted light is almost linearly polarized and the orientation of polarization ellipse is spatially distributed as expected. The intensity distribution of the generated beams exhibits a zero intensity area in the center due to the existence of a polarization singularity [43].

To experimentally confirm our theoretical results we fabricated the dielectric metasurface samples on top of a fused quartz glass substrate by electron beam lithography and a followed plasma etching process (for details see Supplementary Material). The size ($L \times W \times H$) of each nanofin is $180 \times 90 \times 500$ nm$^3$ and the whole metasurface contains $1000 \times 1000$ pixels with a lattice constant of $S=240$ nm. Two scanning electron microscopy images of the fabricated metasurfaces are shown in **Figures 4**a-b. For the experimental characterization of the performance of the metasurfaces we use the setup that is shown in **Figure 4**c. In the experiment a linear polarizer is placed in front of the sample to generate the desired incident linearly polarized beam. Due to the sub-millimeter size of the metasurface, a microscope objective (20×/NA=0.45) is placed behind the sample to magnify and image the generated beam to a charge coupled device camera. For the purpose of characterizing the polarization states of the generated beams by using the Stokes parameter method, another linear polarizer was placed behind the microscope objective.

We calculated the Stokes parameters from the measured intensity in order to examine the spatial variant polarization states of the generated beams. The Stokes parameters can be expressed as follows [44, 45]:

$$\begin{cases} S_0 = I_{0,0} + I_{90,0} \\ S_1 = S_0 \cos 2\chi \cos 2\psi = I_{0,0} - I_{90,0} \\ S_2 = S_0 \cos 2\chi \sin 2\psi = I_{45,0} - I_{135,0} \\ S_3 = S_0 \sin 2\chi = I_{45,90} - I_{135,90} \end{cases} \quad (4)$$

where $\chi$ indicates the ellipticity angle ($-\pi/4 \leq \chi \leq \pi/4$) and $\psi$ ($0 \leq \psi \leq \pi$) represents the orientation angle of polarization ellipse. The $I_{i,j}$ denotes the intensities of the transmitted light after passing through the linear polarizer P2 or the combination of P2 and a quarter-wave plate. That is, the subscript $i$ represents the direction of the transmission axis of P2 and the subscript $j$ indicates an extra phase delay between two orthogonal components of the incident light introduced by the wave plate. By rotating P2, we acquire $I_{0,0}$, $I_{45,0}$, $I_{90,0}$ and $I_{135,0}$. For the purpose of measuring $I_{45,90}$ and $I_{135,90}$, we add a quarter-wave plate in front of P2 with its fast axis is along $x$-axis because the quarter-wave plate can provide the desired extra phase delay to the incident light. The intensity distribution of the measured high-order radial and azimuthal polarization beams are shown in **Figure 5**. Note that the intensity of the generated beams is divided into different segments according to their order P. Meanwhile, by rotating the linear polarizer, the extinct areas will be rotated and this evolution process represents the polarization information of such generated CV beams, which is consistent with our expectation (see the videos S1-S3 in the Supplementary Material). The magnified intensity distributions of the central part of our generated beams can be found in the Supplementary Material. By removing all the polarizers and wave plates from the experimental setup, the measured transmission efficiencies of the three fabricated dielectric metasurfaces samples ($P$=3, 4, 5) are 79.98%, 84.95%, and 84.76% respectively. These values are slightly lower than our simulations results.

Finally, we calculate the ellipticity $X$ and the orientation of the polarization ellipse $\psi$ from the measured intensities (**Figure 6**). We find that the ellipticity of the generated radial and azimuthal polarization beams is ranging from -0.2 to 0.2 while the orientation of the polarization ellipses vary in space in correspondence to their order $P$. Nevertheless, at the same spatial position of the two beams, the orientations of the polarization ellipses differ by 90° as it is expected between azimuthal and radial polarized beams. The residual ellipticity, which can be found by comparing the measured ellipticity (**Figure 6**) to the simulation results (**Figure 3**), may be caused by the fabrication errors and the slight anisotropy of the quartz substrate. More information about the analysis can be found in the Supplementary Materials.

Furthermore, we demonstrate that our method can be used for the polarization encryption of information and potential anti-counterfeiting applications. By using the same principle as for the generation of the CV beams, we achieve an image that possesses spatial inhomogeneous polarization states. For our demonstration we start with a grayscale image with a flower pattern (**Figure 7**a). The grayscale level is assigned to a corresponding polarization angle Φ of the image as noted in **Figure 7**a. Hence, in the final image we expect that each petal and the leaves of the flower have different polarization states. Based on the relation of Φ=2θ as derived from Equation 1, we arrange the nanofins with different orientation angles θ ranging from 0° to 75° with an interval of 15°. The orientation angle of nanofins in unstructured (white) background area of the image is set to 75°. The geometric parameters of each nanofin are the same as for the CV beam generation (**Figure 7**b). The experiment set up for characterizing the polarization of such polarization-encoded image is the same as shown in **Figure 4**c. By rotating the linear polarizer P2 from 0° to 150° while maintaining the polarizer P1 at 0°, different petals and leaves of the flower image will be extinct (**Figure 7**c). For example, when the corresponding polarization angle of some areas of the flower is $θ_1$, such areas will be extinct by rotating P2 to

$\theta_2=\theta_1+90°$. Therefore, the intensity pattern of the flower image changes periodically by continuously rotating the linear polarizer P2 (see the video in the Supplementary Materials). Note that when the beam from the laser is directly propagating through the sample without using any polarizer, we only can capture an ambiguously image with nearly uniform intensity distribution (see **Figure S4** in the Supplementary Materials). However, by placing two linear polarizers on both sides of the sample, we will observe the image and distinguish the spatial inhomogeneous polarization distribution. Hence, the polarization information can be used to encrypt information in such dielectric metasurfaces, which may provide a unique path to achieve applications such as data encryption and anti-counterfeiting. Noted that recently another method was proposed for encoding high-resolution color images with spatial multiplexing into different polarization states of light based on Malus' law [28]. While here we concentrate on the generation of high order vector beams with spatially variant polarization based on rigorous coupled wave analysis and make a thoroughly analysis by using stokes method.

In summary, we have demonstrated a method for spatially modifying the polarization states of light by using a dielectric metasurface with low loss and high efficiency. With our design we can generate high-order radial and azimuthal polarization beams with efficiencies up to 80% in transmission mode. Our proposed method provides an alternative and effective way to generate arbitrary vector beams in compact optical systems which surpassing the traditional counterparts. In addition, by using the same principle we achieved a dynamic image reconfiguration with spatial inhomogeneous polarization states. The technique demonstrates that information can be encoded in spatial variant polarization states and cannot be distinguished in laser beam without extra analysis by using the Stokes parameter method. By cascading such dielectric metasurfaces with Huygens' metalenses, which are composed of nanodisks with polarization insensitive properties, the generated high-order vector beam can show the potential of achieving super-resolution within a very compact and flat optical element.

Meanwhile, it may also be used to tailor the properties of (amplitude, phase and polarization) transmitted light combined with other metasurfaces with different functions. Furthermore, such a method for generating vector beams opens promising avenues for applications in optical trapping, mechanical fabrication processing, polarization encryption, and anti-counterfeiting areas.

**Acknowledgements**

The authors acknowledge the funding provided by the National Natural Science Foundation of China (No. 61775019, 61505007) program and Beijing Municipal Natural Science Foundation (No. 4172057). L.H. acknowledge the support from Beijing Nova Program (No. Z171100001117047) and Young Elite Scientists Sponsorship Program by CAST (No. 2016QNRC001). This project has received funding from the European Research Council (ERC) under the European Union's Horizon 2020 research and innovation programme (ERC grant agreement No. 724306).

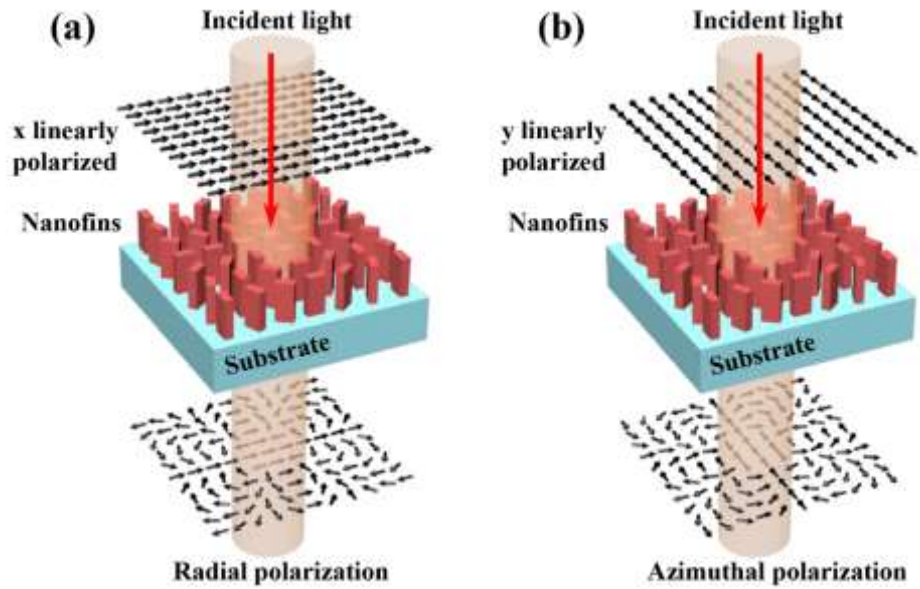

**Figure 1.** Schematic illustrations of high order radial and azimuthal polarization beams (*P*=4) generated by a single dielectric metasurface. The black arrows indicate the spatial inhomogeneous polarization states of light. (a) Radial polarization beam is generated when the incident light is *x* linearly polarized (b) Azimuthal polarization beam is generated when the incident light is changed to *y* linearly polarized.

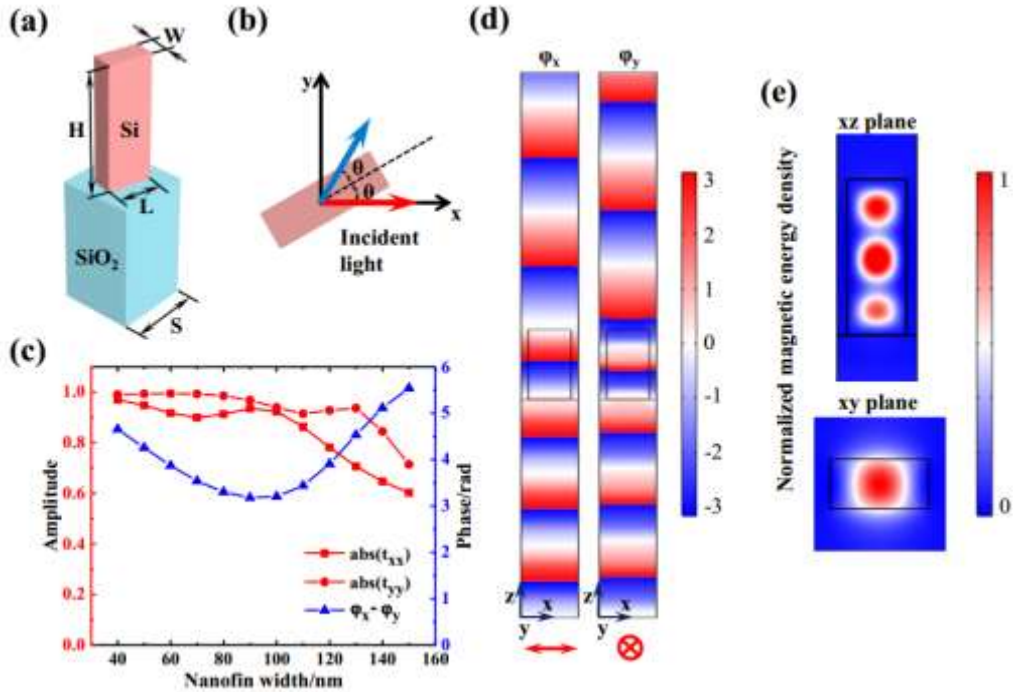

**Figure 2.** (a) Schematic illustration of an amorphous silicon nanofin sets on the glass substrate. (b) Each nanofin acts as a local half-wave plate that can rotate the polarization of the incident light by $2\theta$ while maintaining the transmitted light linearly polarized. The red and blue arrow indicate the polarization directions. (c) Simulated amplitude and phase difference of the transmission coefficients $t_{xx}$ and $t_{yy}$ for a fixed length $L=180$ nm. (d) Simulated phase of $E_x$ with the incident $x$ polarized light and phase of $E_y$ with $y$ linearly polarized light. The light is propagating in the positive $z$-direction. The red arrow and cross indicate the polarization of incident light. (e) Simulated normalized magnetic energy density distribution of a single nanofin ($180\times90\times500$ nm$^3$, $L\times W\times H$) with periodic boundary conditions when illuminating $x$ linearly polarized light with finite element method. The black lines represent the silicon nanofin.

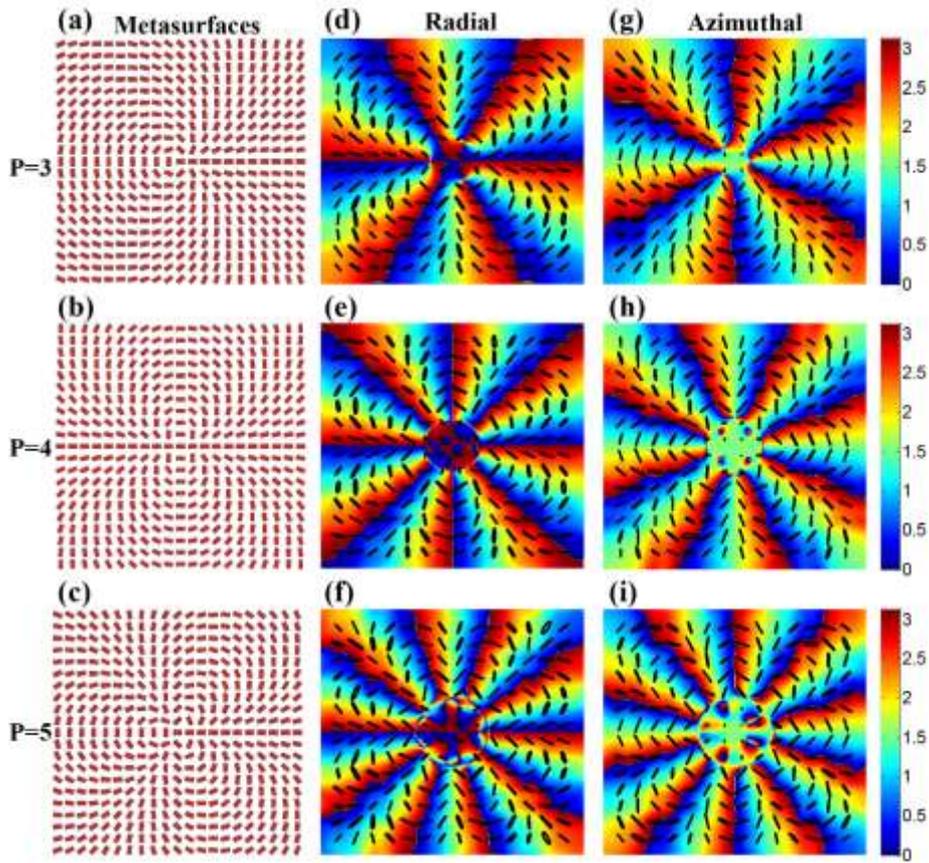

**Figure 3.** (a-c) Arrangements of the nanofins to compose the metasurfaces for generating high-order radial and azimuthal polarization beams (*P*=3, 4, 5). (d-i) Calculated spatial distribution of the polarization ellipse (black ellipse) and orientation angle (color) of the generated beams. The calculation is done using to Equation 3 and the simulation results of electromagnetic field simulation.

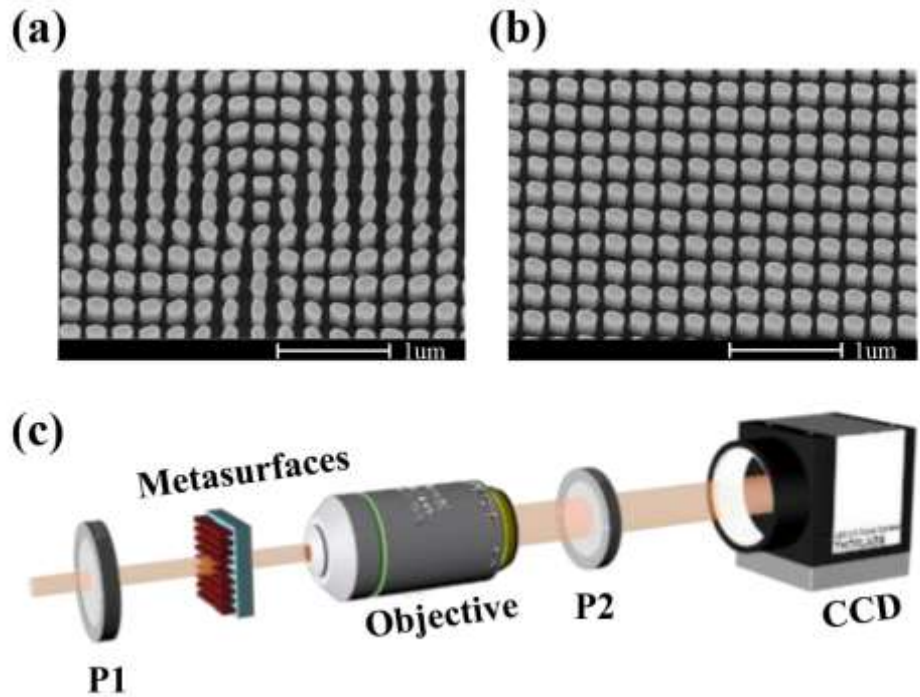

**Figure 4.** (a-b) Scanning electron microscopy images of two fabricated metasurface samples for the generation of high-order radial and azimuthal polarization beams. The geometric parameters of the amorphous silicon nanofins are $180\times90\times500$ nm$^3$. The samples are composed of $1000\times1000$ nanofins with a lattice constant along *x*- and *y*-axis of 240 nm. (c) The experimental setup for generation and observation of the radial and azimuthal polarization beams which consists of two linear polarizers (P1, P2), a 20x magnifying microscope objective and a CCD camera.

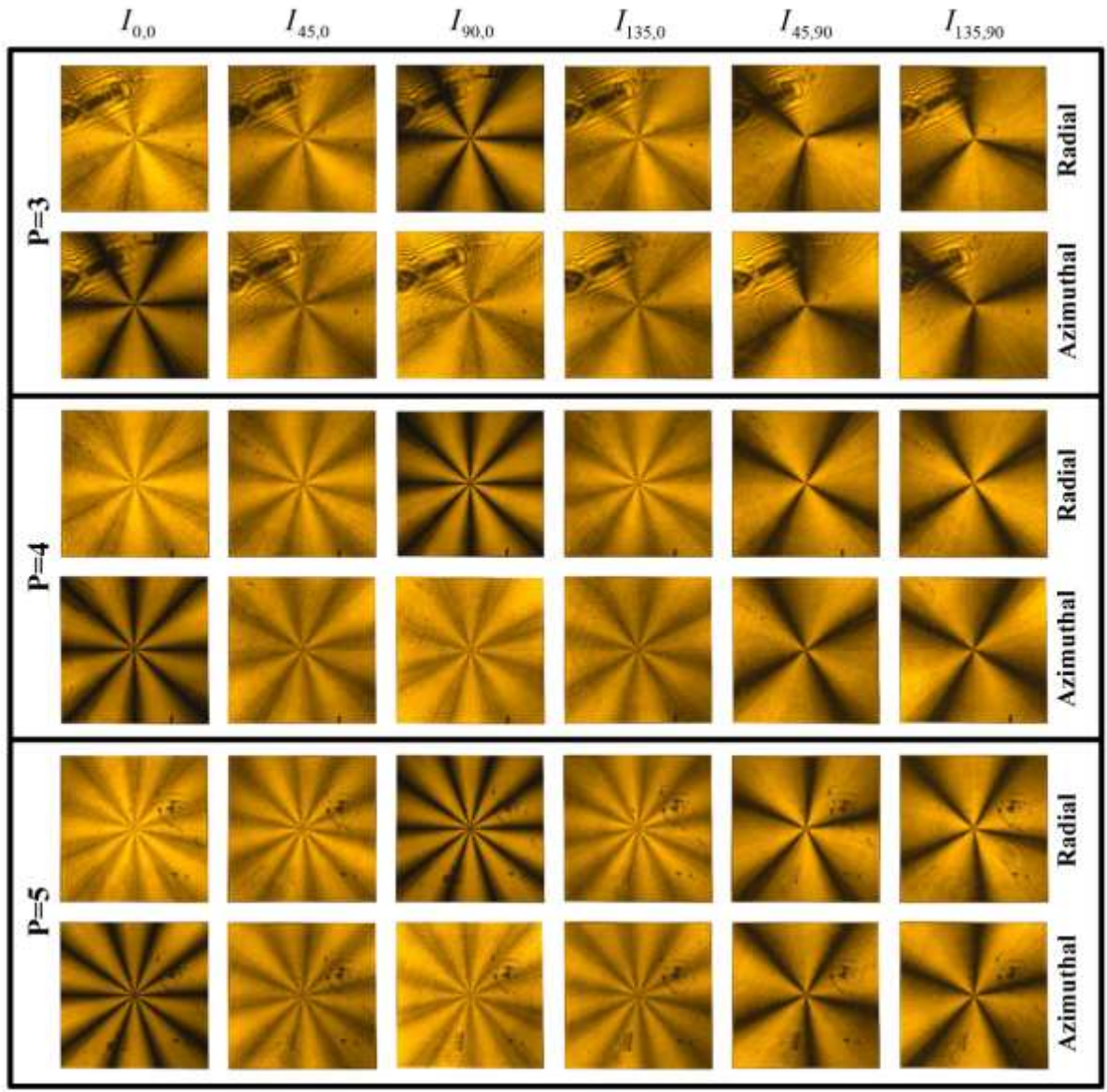

**Figure 5.** The measured intensity distributions of the high-order radial and azimuthal polarization beams for calculating the Stokes parameters. The intensity values $I_{i,j}$ denote the intensity of the transmitted light after passing through the second linear polarizer P2 or the combination of P2 and a quarter-wave plate. By rotating P2, we acquire $I_{0,0}$, $I_{45,0}$, $I_{90,0}$ and $I_{135,0}$.

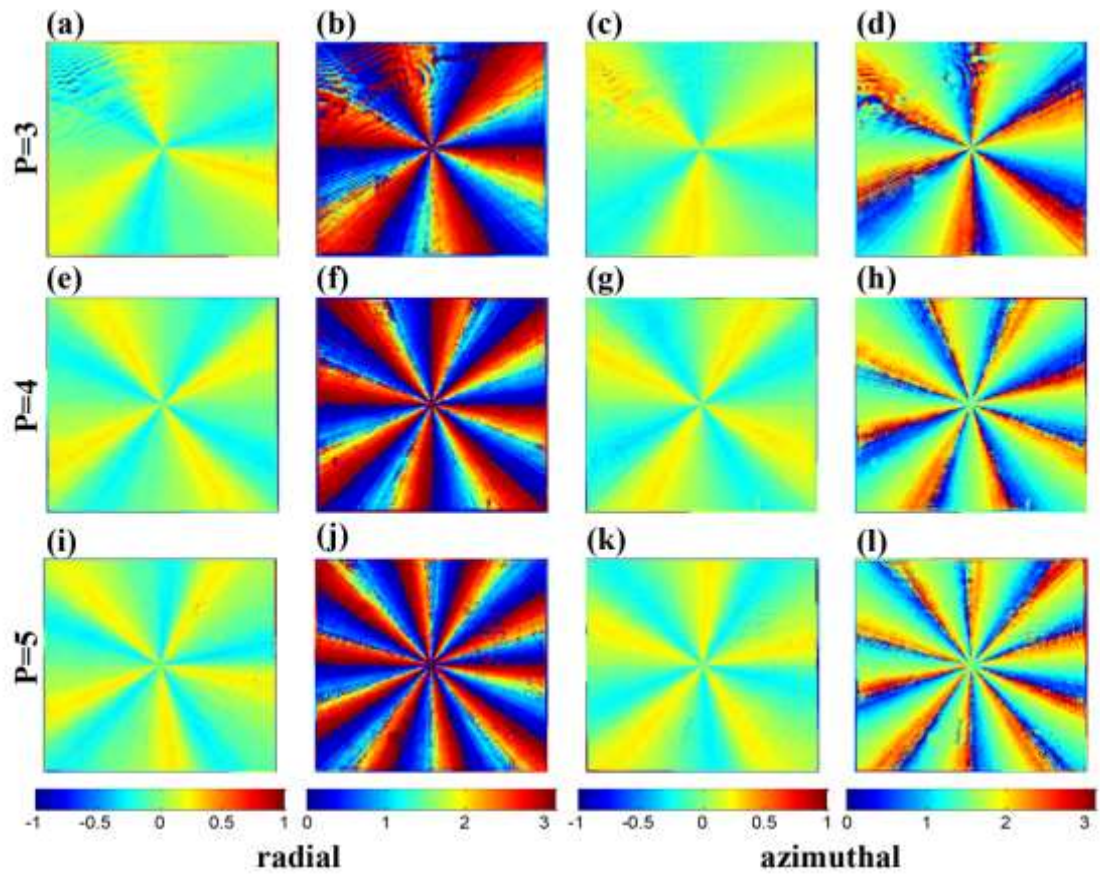

**Figure 6.** The ellipticity and the orientation of the polarization ellipses of our generated high-order (*P*=3, 4, 5) radial and azimuthal polarization beams. They are calculated by Stokes parameters according to Eq. (3) and (4). The radial and azimuthal polarization beams can be generated by *x* and *y* linearly polarized beams by using the same sample at the wavelength of 780 nm.

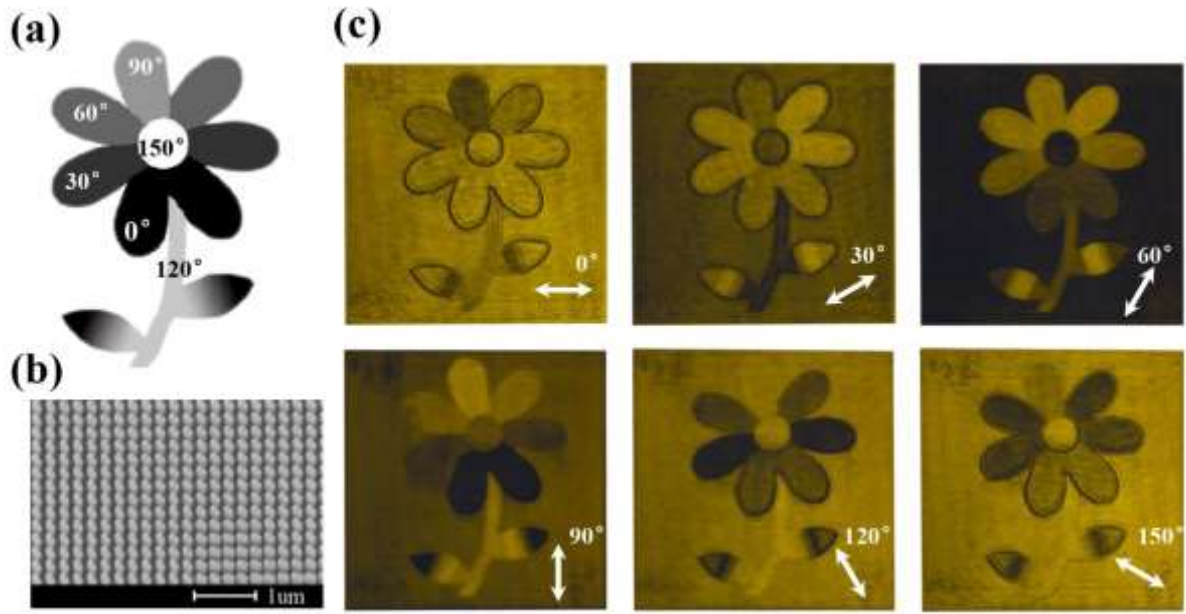

**Figure 7.** (a) Original grayscale image of a flower to achieve dynamic polarization control. The numbers indicate the polarization angle of transmitted light. (b) The scanning electron microscopy (SEM) images of dielectric metasurface samples for experimentally generating the dynamic image. (c) The experiment results of the dynamic image captured by a CCD camera. The white arrows indicate the transmission axis of the linear polarizer P2.